\documentclass{kluwer}    

\begin{document}                                                                                   
\begin{article}
\begin{opening}
\title{Star Formation Histories of Nearby Dwarf Galaxies}
\author{Eva K.\ \surname{Grebel}}  
\runningauthor{Eva K.\ Grebel}
\runningtitle{Nearby Dwarf Galaxies}
\institute{Max Planck Institute for Astronomy, K\"onigstuhl 17, 
D-69117 Heidelberg, Germany}

\begin{abstract}
Properties of nearby dwarf galaxies are briefly discussed.
Dwarf galaxies vary widely in their star formation histories, the ages of their 
subpopulations, and in their enrichment history. 
Furthermore, many dwarf galaxies show evidence for spatial variations in 
their star formation history; often in the form of very extended old 
populations and radial gradients in age and metallicity.  Determining factors
in dwarf galaxy evolution appear to be both galaxy mass and environment.
We may be observing 
continuous evolution from low-mass dwarf irregulars via transition types 
to dwarf spheroidals, whereas other evolutionary transitions seem less likely.
\end{abstract}

\end{opening}           

\section{Introduction}  

Dwarf galaxies are the most frequent type of galaxies in nearby galaxy groups
and clusters, and probably in the Universe.  They are the primary building
blocks of more massive galaxies in hierarchical clustering scenarios and are
believed to have been even more numerous at earlier times.  The best studied
dwarf galaxies are nearby dwarfs where individual stars can be resolved and
evolutionary histories can be derived in great detail.  Hence the most 
detailed information is available for our immediate neighbors, the dwarf
satellites of the Milky Way.  A growing body of data is becoming available 
also for more distant dwarf galaxies in the Local Group and beyond owing 
largely to observations with the {\em Hubble Space Telescope} (HST).  Apart 
from deep observations of selected galaxies our ongoing HST snapshot survey
of nearby galaxies is rapidly increasing the number of dwarf galaxies within
5 Mpc for which resolved upper-red-giant-branch photometry is available.
Furthermore, observations with large 8m -- 10m ground-based telescopes such 
as Keck, Gemini, Subaru, and the Very Large Telescope (VLT) are playing an 
increasingly important role.  

For the
purpose of this review we will consider all galaxies with absolute $V$
magnitudes $M_V > -18$ mag and with diameters of a few kpc or less as dwarf 
galaxies.
Dwarf galaxies are important not only as future constituents of larger
galaxies, but also in their own right.  They span a wide range of masses,
luminosities, mean metallicities, gas content, kinematic properties,
and mean ages.  Dwarf galaxies are found in different environments ranging
from voids to loose groups and dense galaxy clusters, and from relative 
isolation within groups to close proximity of massive galaxies.  Dwarf 
galaxies allow us therefore to study the impact of external environmental
effects and of internal properties such as gas content and galaxy mass 
on galaxy evolution.  Knowledge of their internal kinematics may help to
understand the nature of dark matter, and their global kinematics make them
valuable dynamical probes of the dark halos of massive galaxies and clusters.  

The existing, detailed studies have revealed a surprising diversity in the
star formation histories of dwarf galaxies.  Each galaxy shows evidence 
for a complex evolution that is clearly distinct from the single-age,
single-metallicity history characterizing a typical globular cluster.
Moreover, even within the same morphological subclass no two dwarf galaxies 
are alike and differ considerably in their enrichment histories and/or time 
and length of their star formation episodes.  However, common global 
properties are beginning to emerge, and this review will concentrate on 
identifying them rather than on describing the differences.  For 
recent reviews that describe the properties of Local Group dwarf galaxies 
in great detail we refer to van den Bergh (1999; 2000).   

\section{Types of dwarf galaxies}

A variety of terms are in use for different morphological types of
dwarf galaxies, and different authors use different definitions.  

{\bf Dwarf spirals} comprise S0, Sa, Sb, Sc, and Sd galaxies with
$M_V > -18$, central surface brightnesses of $\mu_V \gsim 23$ mag 
arcsec$^{-2}$, H\,{\sc i} masses of $M_{\rm HI} \lsim 10^9$ M$_{\odot}$, 
large mass-to-light ratios.   Early-type dwarf
spirals are discussed in Schombert et al.\ (1995), while Matthews \&
Gallagher (1997) describe properties of late-type dwarf spirals.  
Dwarf spirals tend to be chemically inhomogeneous and contain a
range of ages just as massive spirals.  Later-type dwarf spirals 
have lower metallicity and less gas (McGaugh 1994) than the earlier types.  
Dwarf spirals may exhibit well-defined spiral structure or may appear to be 
in transition from spirals to irregulars such as Magellanic irregulars (Sm, 
Sdm).  Early-type dwarf spirals show rotation curves typical for 
rotationally supported exponential disks, while late-type dwarf spirals
are slow rotators or exhibit solid-body rotation.  
Dwarf spirals show slow continuous star formation.

NGC\,3109, a galaxy at a distance of 1.33 Mpc in the nearby Sextans-Antlia
group may be considered the closest dwarf spiral since it shows extended
spiral structure (Demers et al.\ 1985) apart from features of an irregular
galaxy.

{\bf Blue compact dwarf galaxies} (BCDs) comprise H\,{\sc ii} galaxies,
blue amorphous galaxies, and certain types of Wolf-Rayet galaxies.  Gas,
stars, and starburst regions tend to be centrally concentrated in BCDs.
Due to their pronounced compact starbursts BCDs have high surface
brightnesses ($\mu_V \lsim 19$ mag arcsec$^{-2}$.  
The H\,{\sc i} masses of BCDs are $\le 10^9$ M$_{\odot}$ and can exceed
the inferred stellar mass.  While BCDs tend to be rotationally supported,
exhibit solid-body rotation and evidence for dark matter, chaotic
motions are detected as well, and part of the extended gas may be 
kinematically decoupled from the galaxies (van Zee et al.\ 1998).  BCDs
may be fitted by $r^{1/4}$ laws in some cases, exponential profiles in 
others,  or composite profiles (Doublier et al.\ 1999). 

The BCD closest to the Local Group is NGC\,6789 at a distance of only 
2.1 Mpc (Drozdovsky \& Tikhonov 2000), while the Local Group does not 
contain galaxies of this type.

{\bf Dwarf irregular galaxies} (dIrrs) are gas-rich galaxies with an
irregular optical appearance usually dominated by scattered
H\,{\sc ii} regions.  They typically have $\mu_V \lsim 23$ mag
arcsec$^{-2}$, $M_{\rm HI} \lsim 10^9 M_{\odot}$,
and $M_{\rm tot} \lsim 10^{10} M_{\odot}$.  The H\,{\sc i} distribution 
is usually clumpy and
much more extended than even the oldest stellar populations.
In low-mass dIrrs gas and stars may exhibit distinct spatial distributions
and different kinematic properties.  Metallicities tend to increase with
decreasing age in the more massive dIrrs, indicative of enrichment.  
Solid body rotation is common, 
though not all dIrrs rotate, especially not very low-mass dIrrs.  
DIrrs are found both in clusters and groups as well as in the field.

The dIrr closest to the Milky Way is the Small Magellanic Cloud (SMC) at
a distance of $\sim 60$ kpc.

{\bf Dwarf elliptical galaxies} (dEs) are spherical or elliptical in 
appearance, tend to be found near massive galaxies,
usually have little or no detectable gas, and tend not to be 
rotationally supported.  DEs are compact galaxies with high central 
stellar densities and are typically fainter than M$_V = -17$ mag, 
have $\mu_V \lsim 21$ mag arcsec$^{-2}$, $M_{\rm HI}
\lsim 10^8 M_{\odot}$, and $M_{tot} \lsim 10^9 M_{\odot}$.
DEs may contain conspicuous nuclei (nucleated dEs, dE(N)) that
may contribute up to 20\% of the galaxy's light.  The fraction of dE,N 
is higher among the more luminous dEs.  
S\'ersic's (1968) generalization of a de Vaucouleurs $r^{1/4}$ law
and exponential profiles describes the surface
density profiles of nucleated and non-nucleated dEs and dSphs best
(Jerjen et al.\  2000a).

The closest dE is NGC\,185, a companion of M31, at a distance of 620 kpc
from the Milky Way.  The closest dE,N is M32\footnote{Note that while
M32 is a {\em dwarf} elliptical according to the definition of a dwarf
galaxy adopted in this paper, but is more akin in its properties to 
classical, giant ellipticals (Wirth \& Gallagher 1984).  Hence NGC\,205
at a distance of about 830 kpc from the Milky Way may instead be 
considered the closest dE,N.}, another M31 companion, which
has a distance of $\sim 770$ kpc from the Galaxy.

{\bf Dwarf spheroidal galaxies} (dSphs) are diffuse, gas-deficient,
low-surface-brightness 
dwarfs with very little central concentration.  They are not always 
distinguished from dEs in the literature.  DSphs are characterized by
M$_V \gsim -14$ mag, $\mu_V \gsim 22$ mag arcsec$^{-2}$, 
$M_{\rm HI} \lsim 10^5 M_{\odot}$, and $M_{tot} \sim 10^7 M_{\odot}$.
They include the optically faintest galaxies known.  DSphs are usually found
in close proximity of massive galaxies
and are generally not supported by rotation. 
Their velocity dispersions indicate the presence of a significant dark
component when virial equilibrium is assumed. 

The closest dSph galaxy is Sagittarius, which is currently merging with 
the Milky Way.  

{\bf Tidal dwarf galaxies} form in mergers and interactions from debris
torn out of more massive parent galaxies.  They do not contain dark matter
and may have high metallicities for their luminosity depending on the 
evolutionary stage of the parent galaxy (Duc \& Mirabel 1998).  Potential
candidates for nearby tidal dwarf galaxies are discussed by Hunter et al.\
(2000).  For more information on properties and formation I refer to the
contributions by Mirabel and by Brinks in these proceedings. 

\section{Star formation characteristics}

Photometric imaging is the method of choice to derive star formation histories
of galaxies that can be resolved into individual stars.  Through comparison
with synthetic color-magnitude diagrams based on evolutionary models very
detailed star formation histories can be determined. 
For more distant objects we have to rely upon integrated colors and spectral 
energy distributions.  More information about models and techniques can be
found in the contributions by Matteucci, by Tosi, and by Bruzual in these 
proceedings.
Metallicities (such as stellar [Fe/H] or nebular oxygen abundances) are 
derived photometrically and through spectroscopy (see contributions
by Peimbert and by Hill in this volume).  The gas content is usually measured
through 21cm observations and narrow-band imaging.   

Individual dwarf galaxies can show a wide range of evolutionary histories
even within the same subclass.  Dwarf galaxies vary widely in the amount of
enrichment that they experienced, in their star formation rates and the 
length of star formation episodes, in their gas content, in their number
of globular clusters (if any), etc.  

\subsection{Old populations} 

A common property of all dwarf galaxies studied in sufficient detail so
far appears to be the presence of an old population, which in many cases
turns out to be the dominant population.  Furthermore, old populations
tend to be spatially more extended than younger ones.  Whether this is
an effect of the increased dispersal of older stars as a function of time,
of expansion due to mass loss, or other effects is unclear.

The term ``old population'' usually refers to stars with ages of 10 Gyr
and more.  These populations can be unambiguously traced through the 
detection of horizontal branch stars or more accurately through the
corresponding main-sequence turnoffs.  Main-sequence turnoffs at the
distance of M31 (770 kpc) occur at $V \sim 28$ mag, which illustrates
why accurate age dating of the oldest populations is impossible in all
but the closest dwarfs.  Horizontal branches are 3 -- 3.5 mag brighter
than the oldest main-sequence turnoffs, but their detection can be
difficult in regions of significant crowding or in galaxies with significant
intermediate-age populations, which can obscure horizontal branches in a 
color-magnitude diagram.    

Deep main-sequence photometry (largely based on {\em HST} imaging) has 
established that a number of Local Group galaxies share a common epoch 
of ancient star formation.  Main-sequence
photometry reveals that the oldest globular clusters in the Milky Way halo
and bulge, in the LMC (an irregular galaxy but not a dwarf according to the
definition adopted here), in the Sagittarius dSph, and in Fornax are 
coeval.  Similarly, the oldest field populations in the dSphs
Sagittarius, Draco, Ursa Minor, Fornax, Sculptor, Carina, and Leo\,II have 
the same relative age as the oldest Galactic globular clusters.

The existence of blue horizontal branches in globular clusters in
M31, in the dIrr WLM, and the dE NGC\,147 
are interpreted as indicative of ages
similar to those of the old Galactic globular clusters.
The blue horizontal branch in the field populations of the dSphs 
Sextans, Leo\,I, Cetus, And\,I, And\,II, and Tucana, in
the dIrr/dSph Phoenix, in the dIrr IC\,1613, and in the dEs
NGC\,185 and NGC\,147 appear to imply comparatively old
ages.  Second-parameter effects other than age, however, may also play an
important role here.  

Possible evidence for delayed formation of the first significant (i.e.,
clearly observable) old population may exist in other dwarf galaxies:
The absence of a blue horizontal branch in the field populations of the dIrrs
SMC, WLM, Leo\,A, and DDO\,210 (as well as in the large spiral M33) may
indicate that the bulk of the old population in these galaxies
formed a few Gyr later than the oldest Milky Way globular clusters.
These galaxies span a range of distances from more massive galaxies,
and there is no obvious reason for the apparent difference in the oldest
significant star formation episodes. 

For a list of references for the studies
of the individual galaxies quoted here see Grebel (2000).   

In dwarf galaxies well beyond the Local Group (at distances of 2 Mpc and
more) the available studies tend not to go deeper than a few magnitudes
below the tip of the red giant branch.  
Both integrated colors and the detection of red giant branches in dwarf
spirals and BCDs indicate the presence of past star formation
episodes in these objects (e.g., Papaderos et al.\ 1996; Lynds et al.\ 
1998; Schulte-Ladbeck in this volume).  Without
photometry at least down to the horizontal branch the age of these
bona fide ``old'' populations is difficult to constrain, but there is clearly
evidence for populations older than 2 Gyr.    

\subsection{Star formation and spatial variations}

Star formation in the disks of dwarf spirals appears to be 
largely driven by spiral density waves.  
Dwarf spirals appear to have experienced
continuous, low-level star formation over a Hubble time and will
likely continue in the same manner for a long time.  Rotation, shear,
metallicity, and H\,{\sc i} surface density tend to decrease toward
later types (e.g., McGaugh 1994; de Blok et al.\ 1995).  

BCDs have one or several centrally concentrated
starburst regions, which may contain super star clusters.
With H\,{\sc i} densities
of up to $\sim 10^{21}$ atoms cm$^{-2}$ in active regions BCDs exceed the
Toomre instability criterion for star formation, which facilitates
their high star formation rate (e.g., Taylor et al.\ 1994, van Zee et
al.\ 1998).  Many
BCDs are observed in isolation without recognizable companions, hence
interactions do not seem to be the agent for the vigorous star formation.

The interstellar medium (ISM) in dIrrs is 
highly inhomogeneous and porous, full of small and large shells and
holes.  Star formation may be driven by homogeneous turbulence, which
creates local densities above the star formation threshold (e.g.,
Stanimirovic et al.\ 1999).  Lower gravitational
pull and the lack of shear in absence of differential rotation imply
that H\,{\sc i} shells may become larger and are long-lived
(Hunter 1997).  Diameters, ages, and expansion velocities of the
H\,{\sc i} shells increase with later Hubble type (Walter \& Brinks 1999)
and scale approximately with the square root of the galaxy luminosity
(Elmegreen et al.\ 1996).  Shell-like
structures, H\,{\sc i} holes, or off-centered gas may be driven by
supernovae and winds from massive stars
following recent star formation episodes or tidal interactions.
Indeed, evidence for outward propagating star formation within a central 
H\,{\sc i} shell was found in the dIrr Sextans\,A in the Sextans-Antlia group
(van Dyk et al.\ 1998).  Numerous
shells with propagating star formation along their rims
were uncovered in the dIrr IC\,2574 in the M81 group (Walter \& Brinks 1999), 
while tidal interactions may be responsible for the off-centered H\,{\sc i}
distribution, asymmetric H\,{\sc i} disks, or counterrotation seen
in the dIrr NGC\,55 in the Sculptor group (Puche et al.\ 1991) and in 
the fairly isolated dIrr NGC\,4449 in the CVn\,I cloud (e.g., Hunter et 
al.\ 1999).  On a global, long-term scale, however, star formation has
essentially occurred continuously at a constant rate with amplitude
variations of 2--3 (Tosi et al.\ 1991, Greggio et al.\ 1993) 
and is largely governed by internal, ``local'' processes (Hunter 1997).  
 
The best-studied dEs are the four dE companions of M31.  They have dominant
old and intermediate populations, but can also show recent, centrally
concentrated star formation as in the case of NGC 185 (Mart\'{\i}nez-Delgado
et al.\ 1999).  The H\,{\sc i} in these dEs ranges from almost non-existent
to counterrotating to being consistent with expectations from stellar mass
loss (e.g., Sage et al.\ 1998).  

DSphs, in contrast, have been found to be devoid of gas within their
optical radii down to column densities of 2 to $6\cdot 10^{17}$ cm$^{-2}$
(e.g., Young 2000).  However, H\,{\sc i} with matching radial velocities
was detected in their surroundings
(Carignan 1999; Blitz \& Robishaw 2000), which may have been removed through
ram pressure effects.  While dSphs have predominantly old and intermediate-age
populations, the intermediate-age fraction increases roughly with 
Galactocentric distance, which may be caused by the decreased impact of
ram pressure and tidal stripping (e.g., van den Bergh 1994).  
Intermediate-age and younger populations, where present, tend to be 
centrally concentrated.  This may indicate that star formation could be
sustained longer in the centers, where gas was retained for a more extended
period.  Even in dSphs that are largely old there is some evidence for spatial
variations in star formation history:  red horizontal branch stars are often
found more centrally concentrated than blue horizontal branch stars (e.g.,
in Sculptor; Hurley-Keller et al.\ 1999).  However, this trend is not
observed in all dSphs --- And\,II is a counterexample (Da Costa et al.\
2000).  The metallicity spread found in ``single-age'' dSphs such as 
the faint Milky Way companions 
Draco und Ursa Minor indicates that their early star
formation episode must have been sufficiently extended to allow for
this enrichment.  While their luminosity functions are indistinguishable
from those of old Galactic globular clusters (Grillmair et al.\ 1998;
Feltzing et al.\ 1999), the abundance ratios of the dSphs suggest that
their nucleosynthetic histories differed from those of
average Galactic halo stars in terms of having lower [$\alpha$/Fe] ratios
(Shetrone et al.\ 2000).   

In summary, the following modes of star formation are observed among 
nearby dwarf galaxies: (1) Continuous star formation with a constant 
or varying star formation rate over a Hubble time and gradual enrichment;
this mode appears to hold for dSphs, massive dIrrs, and possibly BCDs.
(2)  Continuous star formation with decreasing star
formation rate that ceases eventually.  Examples include low-mass dIrrs,
dEs, and most dSphs.  
(3) Distinct star formation episodes separated by Gyr-long periods of
quiescence.  So far only one example of this mode is known, the Carina dSph
(e.g., Hurley-Keller et al.\ 1998).  It is unclear what caused the gaps and the
subsequent onset of star formation, and why this dSph does not show 
chemical enrichment.

Dwarf galaxy evolution as a whole appears to be determined both by 
environmental effects and by galaxy mass.  Indeed
all morphological types of LG dwarf galaxies tend to  follow global
relations between absolute
magnitude, mean metallicity, and central surface brightness.  The more
luminous a galaxy the higher its metallicity.  These relations hold also
for most dwarfs outside of the Local Group.  For a more detailed discussion
see Skillman (these proceedings).   

\section{Potential evolutionary transitions?}

Can dwarf galaxies of one morphological type evolve into another?  

As was argued by van Zee et al.\ (1998), the rotation of BCDs
make it unlikely that these dwarfs could evolve into dEs as they
would need to get rid of their angular momentum.  Also, BCDs are
often found in the field, whereas dEs are predominantly
found in dense cluster environments.  Similarly, 
the compact, concentrated structure of BCDs suggests they do not 
evolve into dIrrs.  Nor is there an obvious mechanism to achieve the
required expansion as dIrrs have a by a factor 2 larger
envelope scale length.  Under favorable conditions, evolution
from BCDs to {\em nucleated} dwarf ellipticals may be possible (Marlowe
et al.\ 1999). 

An interesting case combining both properties of a dE and a
spiral was recently uncovered by Jerjen et al.\ (2000b):  
They show that the dE,N
IC\,3328 shows weak underlying spiral structure and is likely a nearly
face-on dS0 galaxy.  Knezek et al.\ (1999) studied three mixed-morphology,
gas-rich transition-type candidates and found that neither of them is
likely to evolve into a dE over the next Hubble time. 

The presence of intermediate-age or even young populations in some of the
more distant dSphs, the possible detection of associated gas in the
surroundings of several dSphs, indications of substantial mass loss,
morphological segregation, common
trends in relations between their integrated properties, and the
apparent correlation between star formation histories and Galactocentric
distance all seem to support the idea that low-mass dIrrs may eventually
evolve into dSphs, which may be fostered by external effects such as
ram pressure and tidal stripping.  
The dSph Fornax with its significant young (100--200 Myr) population 
despite the absence of gas may represent an advanced stage of such a
transition (Grebel \& Stetson 1999).

The [O/Fe] abundances in dSphs were found to be systematically higher
than in other galaxies, particularly dIrrs (Richer et al.\ 1998).  The
ratio of [O/Fe] serves as a measure of
the star formation time scale, since Fe is produced by SNe Ia and II with a
significantly longer enrichment time scale than O.  As
dSphs lack H\,{\sc ii} regions, direct measurements of their O 
abundances are based on planetary nebulae.  Planetary nebulae were 
detected in only two dSphs so far, namely Fornax and Sagittarius, which
are also the two most massive dSphs.  [O/Fe] ratios in dIrrs, on the
other hand, are derived from H\,{\sc ii} region abundances.  As discussed
in Richer et al.\ (1998), these measurements represent the maximum of the 
stellar O abundances, whereas planetary nebulae are a measure of the
mean stellar O abundance.  A correction for this increases the 
difference in [O/Fe] ratios in these two types of galaxies further.
However, I am not aware of similar, published [O/Fe] ratio measurements
in dIrrs of comparable mass as dSphs, i.e., with masses of a  few times
$10^7$ M$_{\odot}$.  Such low-mass dIrrs include LGS\,3, Phoenix, and 
GR\,8, and are also called ``transition-type'' galaxies to indicate that
they may be evolving from dIrrs to dSphs.  It is important to compare
dSphs to this specific type of low-mass dIrr since only here galactic 
wind properties and galactic potential
(which determine chemical enrichment) may have 
been comparable to those in dSphs if we assume that dSphs are not the
result of a catastrophic event such as a merger, nor underwent extreme
mass loss.  Similarity is furthermore supported by the fact that the
measured stellar velocity dispersion of
LGS\,3 is comparable to that of dSphs (Cook et al.\ 1998).
Mateo (1998) argues that the three transition-type galaxies lie on
the same branch as dSphs when plotting [O/H] or [Fe/H] versus 
absolute magnitude.  Thus the chemical properties of dSphs do not seem to
contradict the proposed evolution from low-mass dIrrs to dSphs
outlined above, though additional data
would certainly be useful.
The distinction between dSphs and dIrrs may be more a matter of 
semantics than of physics. 

\acknowledgements
I thank the organizers for inviting me for this keynote talk and for their
financial support.  Many thanks also to Grazyna Stasinska and Jay Gallagher
for their comments on this paper.

\end{article}
\end{document}